**Multidimensional nanoarchitectures for improved indoor light harvesting in dye-sensitized solar cells**

Javier Castillo-Seoane,[a] Lidia Contreras-Bernal,[a,*] Antonio J. Riquelme,[b,*] Samuel Fauvel,[b] Yann Kervella,[b] Jorge Gil-Rostra,[a] Gabriel Lozano,[c] Angel Barranco,[a] Renaud Demadrille,[b] Juan R. Sánchez-Valencia[a] and Ana Borrás[a]

[a]*Nanotechnology on Surfaces and Plasma Lab., Materials Science Institute of Seville, ICMS, CSIC – US, Avd. Américo Vespucio 49, 41092, Seville (Spain)*

[b]*Université Grenoble Alpes, CEA, CNRS, INP, IRIG-SyMMES, F-38000 Grenoble, France.*

[c]*Multifunctional Optical Materials, Materials Science Institute of Seville, ICMS, CSIC – US, Avd. Américo Vespucio 49, 41092, Seville (Spain)*

Keywords: dye-sensitized solar cells, indoor energy harverster, 1D-nanostructured electrodes, plasma and vacuum tecniques, light scattering.

Dye Sensitized Solar Cells (DSSCs) have recently regained attention for indoor light harvesting and powering wireless devices. Advantages such as cost-effectiveness, flexibility, wide angular response, and lightweight design have driven the fostering of the implementation of advanced photonic architectures, dedicated photosensitizers and compatibility with wearable carriers. However, to fully exploit their potential, crucial aspects require further attention, in particular the improvement of spectral compatibility and low-light harvesting mechanisms, as well as the development of efficient photoanodes through high-yield scalable methods. In this article, we propose the use of nanocomposite photoanodes integrating mesoporous $TiO_2$ (m-$TiO_2$) nanoparticles, ITO nanotubes (NTs) and $TiO_2$ anatase shells (ITO@$TiO_2$ NTs) prepared by step-by-step method relying on mild temperature conditions and avoiding toxic precursors. These photoanodes outperform previous attempts to implement low-dimensional ITO and ITO@$TiO_2$ nanowires and nanotubes for outdoor light conversion, demonstrating an outstanding PCE under low artificial light intensity of 24 % for at 0.014 mW·cm-2, a 166 % increase compared to the conventional architectures. Advanced microstructural, optical, and electrochemical characterizations have revealed that the strong scattering effect of the light in the visible range coupled with enhanced charge collection at low-intensity illumination are the essential mechanisms responsible for such enhanced energy conversion. Remarkably, our devices retain up to 90% of the normal incidence efficiency even under glancing illumination, while conventional reference devices show a drop down to 50%.

1. Introduction

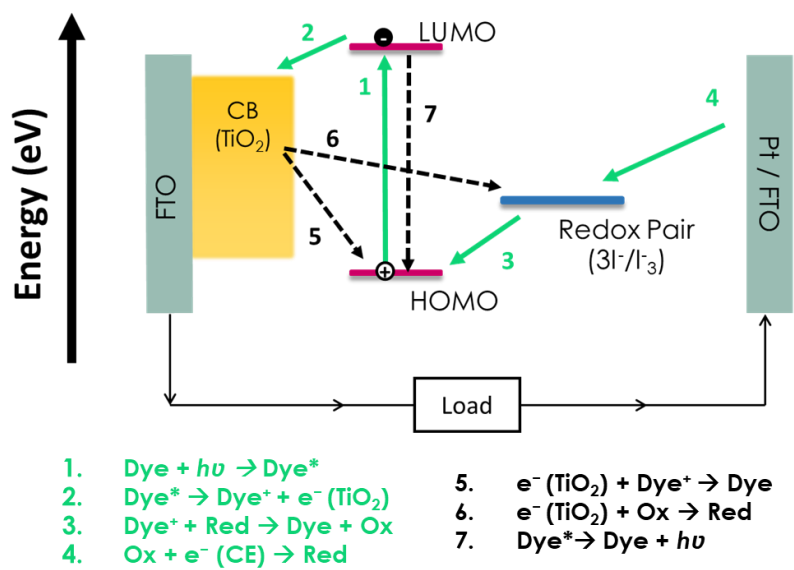

**Scheme 1.** Simplified scheme of the electronic bands and main recombination mechanisms of dye sensitized solar cell. Solid and dashed arrows represent processes in favour or against the charge collection, respectively. 1: Excitation of electrons from the dye molecule (Dye*) by photons; 2: Electrons injection into conduction band (CB) of the mesoporous $TiO_2$ layer leaving the dye molecule in an oxidized state ($Dye^+$); 3: Dye regeneration by receiving electrons from reducing species (Red, $I^-$) of the electrolyte; 4: Oxidizing species (Ox, $I^-_3$) regeneration of the electrolyte by receiving electrons from platinum counter-electrode; 5: Injected electrons into CB of $TiO_2$ layer back into the dye; 6: Injected electrons into CB of $TiO_2$ layer back into the electrolyte; 7: Radiative recombination from LUMO to HOMO of the dye.

Harnessing light energy in indoor environments holds great promise for advancing the Internet of Things (IoT) by powering low-consumption (microwatts level) and portable electronic devices.[1,2] Among the various photovoltaic approaches, dye-sensitized solar cells (DSSCs) have re-emerged as key players due to their affordability, robustness and durability, aesthetic appearance including multi-colour and semitransparency,[3–5] and direct integration with light management systems.[6–8] Such characteristics also make DSSC technology attractive for its implementation in building integrated photovoltaics (BIPV) and dim light harvesting.[9,10] In recent years, progress has been made to improve the power conversion efficiency (PCE) under low-intensity light and also under illumination from diffuse or glancing light sources.[11–13] This progress has been facilitated by the development of specialized photosensitizers designed to optimize spectral compatibility with artificial light sources.[14,15] In addition, a new generation of electrodes, including both photoanodes and counter-electrodes, has been introduced to improve charge transport within the redox electrolyte. This advance addresses the most limiting mechanism affecting the photovoltaic performance of dye-sensitized solar cells operating under low-intensity light conditions.[16] Under such conditions, the number of excited dye molecules is reduced and, therefore, the effect of the electron recombination with the oxidized species in both the electrolyte and/or the oxidized dye molecules (reaction mechanisms 5 and 6 in **Scheme 1**) can seriously affect the PV response of the device.[17] In other words, the kinetic competition of the dye regeneration by the the electrolyte redox reaction and the charge recombination is more pronounced under indoor lighting conditions than under outdoor lighting (see Scheme 1).[18] It is therefore necessary to regulate the number of redox species of the electrolyte to minimize the recombination processes at low-intensity conditions. This problem, together with the effect of light intensity and its corrosive nature, has led to the replacement of standard iodide/triiodide electrolytes with alternatives based on cobalt and copper. In this context, Freitag *et al*. have recently published an unprecedented PCE of 38% under fluorescent lighting at 1000 lux using a copper(II/I) electrolyte.[19] Achieving higher efficiency therefore depends on improving the absorption of incident light, and optimizing charge carrier separation, and transport. These factors are closely linked to the nanostructure of the photo-anode. Among the preferred photo-anode materials, $TiO_2$ mesoporous oxide films have gained widespread use due to their potential in both indoor and outdoor applications. The internal surface area and porosity of the mesoporous film have a significant effect on dye chemisorption, which determines the photocurrent generation. In addition, the morphology and structure of the photo-anode have a critical effect on the charge transfer kinetics. To overcome the problem of slow electron diffusion across $TiO_2$ nanoparticles boundaries, researchers have turned to one-dimensional (1D) nanostructured materials such as nanotubes (NT), nanowires (NW), and nanorods (NR) as a substitute for nanoparticles.[20] These 1D nanomaterials provide directional electron transport pathways and reduce charge recombination due to their low grain boundary defects.[20–24] In addition, one-dimensional arrays typically exhibit a more pronounced light-scattering effect compared to mesoporous layers, thereby increasing light-harvesting efficiency. Taking such considerations into account, 1D nanoelectrodes emerge as promising candidates under low light intensities by increasing the number of excited dye molecules and thus reducing recombination mechanisms. However, one of the main limitations of 1D nanomaterials is their inherently low surface area in compared to mesoporous nanoparticles.[25] To overcome this drawback, we propose here an alternative structure for the photoelectrode of DSSCs that integrates low-dimensional core@shell nanotubes and mesoporous $TiO_2$ layers. These core@shell nanotubes combine transparent conductive oxide (TCO) - specifically indium-doped tin oxide (ITO) - with a thin conformal layer of highly texturized anatase $TiO_2$. Core@shell nanostructures offer several advantages, by facilitating electron transport and suppressing electron recombination. In the case of ITO@$TiO_2$ nanostructures, the electric field created between the layers enhances the injection of electrons from the $TiO_2$ to the ITO during DSSC operating, reducing electron recombination from the $TiO_2$ to the electrolyte.[26] Although the implementation of low-dimensional semiconducting nanostructures (mostly ZnO and $TiO_2$ NW, NR and NT) has been extensively reported for both dye-sensitized and organometallic halide perovskite solar cells, the heteroarchitecture proposed here has hardly been attempted so far,[21] and moreover its implementation in DSSCs operating at low light intensity.

In this article, core@shell nanotubes are prepared using a fully scalable vacuum and plasma-based technology that is easily extendable to alternative combinations of TCOs and semiconducting oxides and is compatible with large area and roll-to-roll processing. The nanotubes are formed using a soft-template process that uses single crystalline organic nanowires as vacuum-processable 1D templates.[27] This method offers compatibility with a wide range of processable substrates, including organic and polymeric supports. It is carried out at mild temperatures, requires low power consumption, is solvent-free and avoids corrosive precursors... The method has been successfully applied to the development of hybrid, organic, semiconducting, metallic, and piezoelectric nanowires, nanotubes, and nanotrees, with amorphous or highly textued crystalline structures. The nanowires or nanotubes grow according to a vertical open porous arrangement which allows the infusion of the $TiO_2$ nanoparticles in intimate contact with the core@shell nanotubes, increasing the dye absorption surface area and promoting electrolyte diffusion while maintaining intimate contact with the inner transparent and conducting shells. We will also discuss how the size and orientation of the nanotubes promote the scattering of transmitted and reflected light increasing the light absorption capacity of the nanocoatings. The optical and transport properties of such nano-architectures increase the PCE under low artificial light intensity of the resulting DSSC, reaching 24 % at 0.014 mW·$cm^{-2}$, a 166 % increase over the corresponding mesoporous DSSC. Indeed, our 1D nanoelectrodes lead to DSSCs with higher PV response at indoor illumination sources than those DSSCs prepared using commercial photo-anode with a thicker mesoporous layer. Finally, we will analyse the role of the light scattering effects on the performance of the cells under artificial light operation at different angles of incidence.

2. Results and discussion

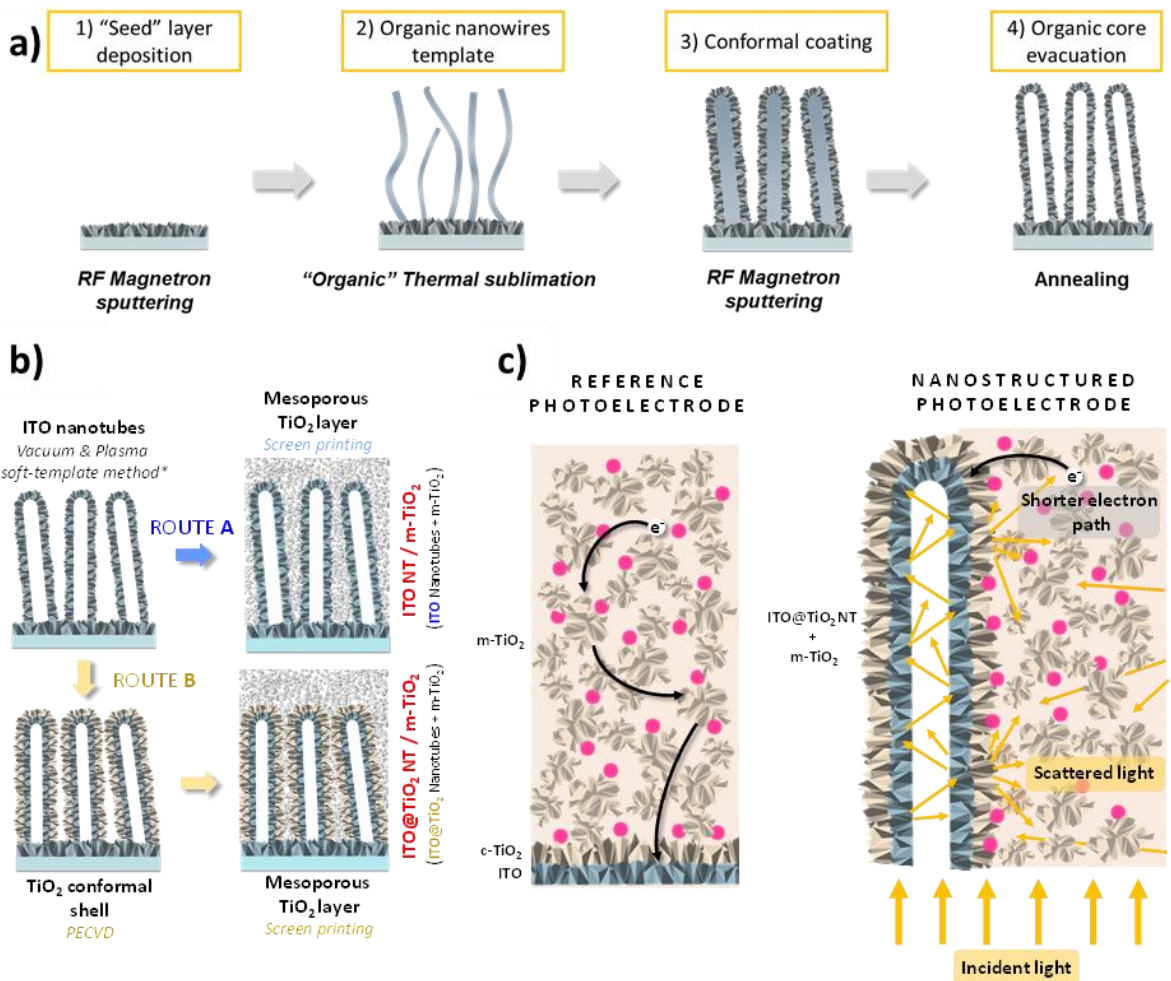

**Scheme 2.** Schematic representation of: a) the manufacturing procecess ITO nanotubes by vacuum and plasma soft-template, b) the manufacturing process of nanostructured photoelectrodes of ITO and ITO@TiO$_2$ nanotubes embedded in the *m*-TiO$_2$ layer, and c) electron transport mechanism in the conventional and nanostructured photoelectrodes. *Vacuum and plasma multi soft-template method described in reference [27].

## 2.1 Fabrication of core@shell nanostructures. Microstructural and optical characterization

The synthesis of the nanocomposite photoelectrodes was approached through a step-by-step procedure based on the combination of organic nanowire (ONW) soft-template techniques (see **Scheme 2**a) enabled by the plasma single-reactor deposition of functional shells and screen-printing of mesoporous nanoparticles (Scheme 2b). The main steps of the method can be described as: 1) fabrication of the ITO thin film (~150 nm) "seed" layer by magnetron sputtering on a commercially available ITO-glass substrate to increase the density of nucleation sites on the substrate; 2) physical vapour deposition of free metal phthalocyanine organic nanowires on the as-grown ITO surface; 3) conformal deposition of a second conductive ITO shell on the organic nanowires used as template. In the final step to form the ITO NTs array, the organic template residues are removed by annealing at 350 °C under air conditions. After the annealing step, the NTs appear partially vertically aligned with lengths between 3 µm and 7 µm, with an average width of 142 ± 19 nm (see Figure 1a and 1c). Figure S1 in supporting information shows the statistical analysis of the NT widths.

For the synthesis of core@shell NTs, a $TiO_2$ film is grown on the ITO NTs by plasma enhanced chemical vapour deposition (PECVD) as it is shown in route B of Scheme 2.[20,28,29] The experimental conditions reported in reference [30] (see also Experimental section) were chosen to ensure the formation of the anatase phase at relatively mild temperatures (below 300 ºC). The SEM images in **Figure 1**b and 1d present the core@shell ITO@$TiO_2$ with an average width of 350 nm (i.e., the thickness of the $TiO_2$ shell is ~95 nm, see also Figure S1).

The thickness of the shell is confirmed by the TEM image as well as by the EDX profile of a single ITO@$TiO_2$ NTs (Figure 2a and 2c). These results also support the high conformality of the $TiO_2$ shell growth, which exhibits a columnar structure characteristic of highly textured anatase layers.[20] The interplanar distances of the shell crystals are consistent with the anatase crystallinity (Figure 2b). These results are in agreement with the XRD analysis carried out on ITO and ITO@$TiO_2$ NT samples. In both cases, the formation of the ITO crystalline structure is evidenced by the presence of (2 1 1), (2 2 2) and (4 0 0) diffraction peaks (blue squares in Figure 2d). Moreover, the main anatase diffraction peak at 25.4º (orange triangle) is also observed for the ITO@$TiO_2$ samples despite the low shell thickness and the lower diffraction capacity of $TiO_2$ compared to ITO. To facilitate the observation of this small peak, a zoom of the corresponding area is shown as an inset in the lower left of Figure 2d).

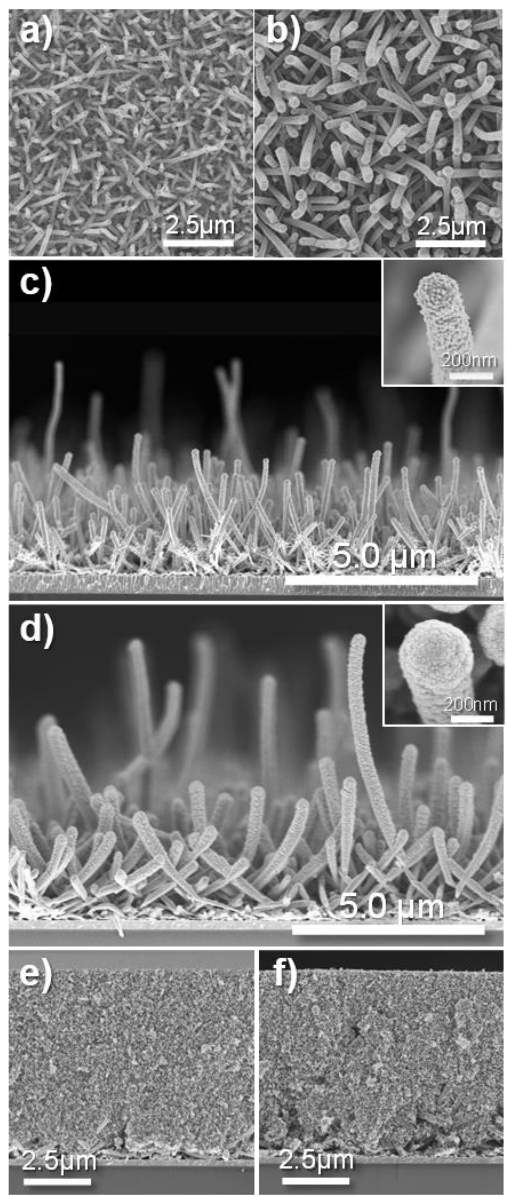

**Figure 1**. Normal-view SEM images of (a) ITO and (b) ITO@TiO2 nanotubes. Cross-sectional SEM images of ITO NTs (c), ITO@TiO$_2$ NTs (d), ITO NTs + m-TiO$_2$ (e) and ITO@TiO2 NTs + m-TiO$_2$ (f).

As mentioned above, the 1D nanoelectrodes have been combined with TiO$_2$ nanoparticles (*m*-TiO$_2$) to form hetero-structured photoelectrodes with enhanced dye absorption. Thus, the final step in the synthesis of the nanocomposite photoelectrodes was the screen-printing deposition of TiO$_2$ nanoparticles (20 nm of nanoparticle size) (routes A and B of Scheme2b for ITO NT and ITO@TiO$_2$ NT, respectively).

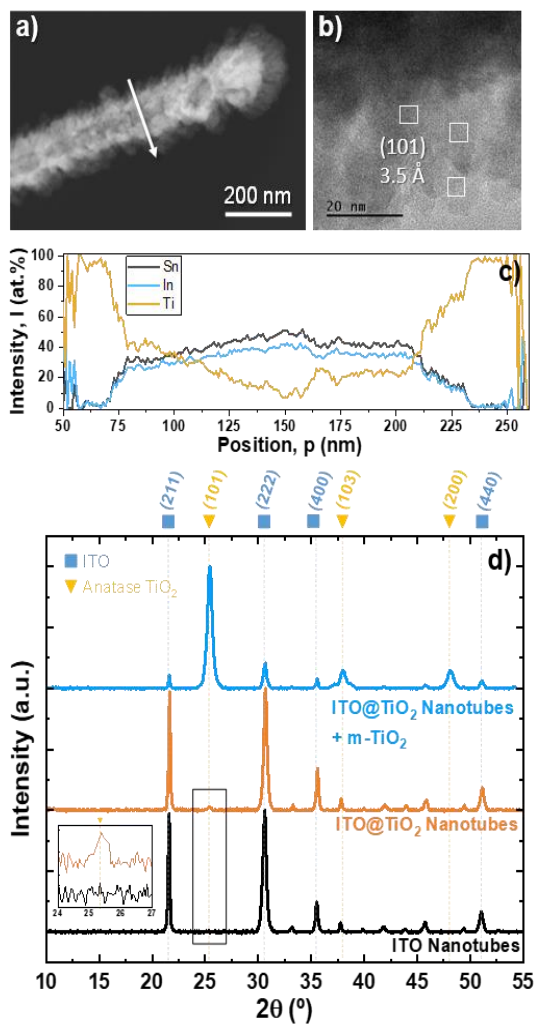

**Figure 2**. a) Representative TEM micrograph of ITO@TiO$_2$ nanotube. b) HRTEM image of TiO$_2$ shell. c) EDXS profile of the ITO@TiO$_2$ nanotube cross-section. d) XRD indexed diffractograms of ITO and ITO@TiO$_2$ nanotubes, and ITO@TiO$_2$ nanotubes embedded into a mesoporous scaffold of TiO$_2$.

Specifically, a total thickness of 8 μm was deposited to ensure a homogeneous coverage of the 1D nanoelectrodes. The cross-sectional SEM images in Figure 1e-f) show the complete coverage of the 1D nanostructures (ITO and ITO@TiO2 NT in Figure 1e and 1f, respectively) embedded in the *m*-TiO$_2$ layer which forms a matrix occupying all the free space between the nanotubes with a slight decrease in the density of the nanoparticles at the interface with the substrate. Figure S2 shows a detailed cross-section view obtained by Focused Ion Beam (FIB), where it is possible to observe the close contact between the *m*-TiO$_2$ nanoparticles and the supported nanotubes. Note that the XRD pattern corresponding to the ITO@TiO2 NT sample (blue curve in Figure 2d), although dominated by the (101) peak of the anatase phase for the *m*-TiO$_2$, shows the peaks corresponding to the ITO.

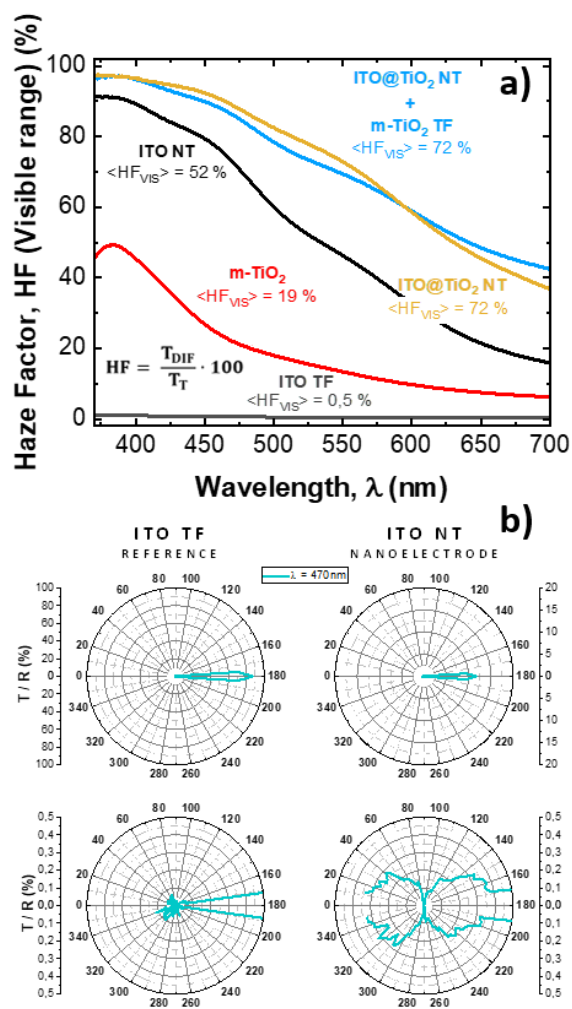

**Figure 3.** a) Haze factor obtained from the transmittance UV-Vis spectra for ITO@TiO$_2$ nanotubes, ITO@TiO$_2$ NT + *m*-TiO$_2$ and thin film *m*-TiO$_2$ reference samples. b) Top) Angular distribution of the light reflected and transmitted of an ITO thin film reference sample and an ITO nanotube sample for wavelength of 470 nm. Bottom) Zoom-in of the angular distribution for reflectance and transmittance showing higher scattering effect on the nanostructure surface.

Figure S3 Left summarizes the optical analysis of the ITO NTs (top), ITO@TiO$_2$ NTs (middle) and ITO@TiO$_2$ NTs/$m$-TiO$_2$ (bottom) samples, including transmittance, reflectance and absorbance in the UV-vis-NIR region for total and diffuse components. Focusing on the visible range (400 nm < λ < 700 nm), significant light scattering is observed even for the sample with infused nanoparticles. Indeed, the diffuse component coincides with the total transmittance in a large part of the visible range for all three cases. This coincidence of the diffuse and total transmittance leads to high Haze factors in the visible range, as shown in Figure 3a.[30] It is worth noting that the diffuse contribution is much more pronounced for both ITO@TiO$_2$ NTs and ITO@TiO$_2$ NTs/$m$-TiO$_2$ than for electrodes of the reference without 1D nanostructures (here called $m$-TiO$_2$ reference sample), and null for ITO TF sample. In addition, the total reflectance of the ITO NT and ITO@TiO$_2$ NT is almost identical to the diffuse component in the visible region. To further demonstrate the strong dispersive character of the nanostructured electrodes, the angular distribution of the light transmittance and reflectance components of an ITO NT sample is also compared with that of ITO thin film (ITO TF) sample. The results in Figure 3b show a decrease in the ballistic transmittance at short wavelength (λ= 450 nm, equivalent angular distribution for 750 and 1000 nm in Figure S3 Right)) for the sample ITO NT sample (see zoom in in the bottom panels), while the ballistic component remains almost unchanged in the case of the sample ITO TF. The effect of such scattering effect on the performance of DSSCs will be analysed in the following sections, considering that most of the dyes for the DSSCs have the maximum light absorption at short wavelengths in the visible region of the electromagnetic spectrum.[31]

## 2.2 Nanotubes of ITO and ITO@TiO$_2$ for dye sensitized solar cells

Transparent Indium Tin Oxide electrodes are one of the most widely used in third-generation solar devices such as perovskite, organic, or dye-sensitized solar cells.[32,33] As mentioned above, the introduction of 1D nanostructured electrodes into photovoltaic devices offers improved properties and functionalities. This section describes the impact of the nanocomposite electrodes as photoanodes in DSSCs, with particular emphasis on indoor and low light performance. In particular, we have focused on two photoelectrodes: the ITO NTs and the ITO@TiO$_2$ NTs both embedded in the *m*-TiO$_2$ layer to enhance dye absorption (hereafter referred to as ITO NT/*m*-TiO$_2$ and ITO@TiO$_2$ NT/*m*-TiO$_2$, respectively). These electrodes correspond to the micrographs in Figure 1e-f. As a reference, we have also used commercially available ITO glass substrates coated with the same anatase TiO$_2$ mesoporous framework as the 1D nanostructured samples. The nanostructured electrodes (and the reference samples) were sensitised with the organic dye YKP-88, using a co-absorbent (chenodeoxycholic acid) in a ratio of 1:1 ratio. YKP-88 was chosen for its good performance and stability (chemical structure in Figure S4).[34] Figure S5 shows the absorptance specta of ITO@TiO$_2$ NT/*m*-TiO$_2$ electrode and reference samples sensitised with the dye. Although both spectra appear saturated at short wavelength, it can be seen that the corresponding to the NT photoelectrode presents stronger absorptance in all the visible range. To complete the DSSC, we used a platinum counter electrode on FTO glass (fluorine tin oxide) and an electrolyte based on the iodide/tri-iodide redox pair in acetonitrile as previously used. [35]

**2.3 DSSCs based on ITO and ITO@TiO$_2$ NT for indoor lighting**

DSSC is one of the most promising technologies for wireless powering the IoT and wireless sensor networks (WSN). One of the main reasons for this is its good performance in low light-intensity conditions due to its high open-circuit voltage for diffuse and low-intensity light conditions, as well as its wide angular compared to other PV technologies. [11,36–38]

This section describes the effect of ITO NT/*m*-TiO$_2$ and ITO@TiO$_2$ NT/*m*-TiO$_2$ electrodes on the PV performance of DSSCs operating in indoor environments with artificial light. We also analyse the effect of the angle of incidence of light on the PCE of these devices. It is worth noting that, for this study, we have obtained the *JV* curves of DSSCs at a wide range of low light intensities following the recommended light intensities for typical real world such as offices or schools (~0.1 mW/cm$^2$) as well as home lighting (0.04 - 0.014 mW/cm$^2$) (see Figure S6 for the emission spectra of the tested light sources).[39–42] **Figure 4**a shows the *JV* curves of the champion DSSCs obtained under this variety of illumination intensities (0.1, 0.04 and 0.014 mW/cm$^2$) from a cool-white LED of 6500K (hereafter called Lamp 1, see spectra in Figure S6) while **Table 1**, Figures 5b, and S7 show the corresponding PV parameters statistics of each type of DSSCs and illumination intensity using Lamp 1. An increase in the PV efficiency of the 1D nanostructure-based DSSCs is clearly observed as the light intensity decreases (see Figure 4b). Here, the TiO$_2$ shell provides better $V_{oc}$ than the ITO NT/m-TiO$_2$ but at the cost of losing $J_{sc}$. Indeed, at 0.014 mW/cm$^2$ the highest PCE is obtained for the ITO NT/m-TiO$_2$ -based DSSC, reaching 22.6 %, followed by the ITO@TiO$_2$ NT/m-TiO$_2$ and the reference samples (efficiencies of 15.2 % and 8.5 %, respectively). The positive effect of 1D ITO nanostructures becomes even more evident when comparing their PV performance with that of a DSSC containing a commercial photoanode (from *Solaronix*) (Figure 4b). It is noteworthy that although the *Solaronix* sample contains a thicker mesoporous layer of TiO$_2$ on FTO glass (13 µm), it does not exceed the efficiency of ITO NT/*m*-TiO$_2$- based DSSC in the light intensity range of 0.04-0.014 mW/cm$^2$.

Figure 4 c) shows the normalized efficiencies obtained from *JV* curves measured at different angles of illumination, from normal to glancing, using the indoor lamp 1. From this experiment, it can be seen that the efficiency of the reference samples decreases by 70 % when comparing normal incident illumination (note that the Figure 4 c) is normalized to this PCE at 0º) with an almost vertical position with respect to the light source. Surprisingly, ITO NT/*m*-TiO$_2$ and ITO@TiO$_2$ NT/*m*-TiO$_2$-based DSSCs retain 90% and 80% of efficiency respectively, under the same measurement conditions. This result highlights the positive impact that the incorporation of ITO and ITO@TiO$_2$ NTs in the photoanodes has on the PV performance of DSSCs operating at low-light intensities and under realistic light directions.

The effect of different interior lamps on the PCE of the DSSCs is also studied in Figure S8. Specifically, we have recorded the *JV* curves of the DSSCs using warm-white 2700K LEDs with different luminous fluxes (Lamp 2 and Lamp 3), and a cool-white fluorescent lamp (Lamp 4) as illumination sources (see the emission spectra in Figure S6). By adjusting the light intensity to 0.04 mW/cm$^2$ for all the illumination sources and the distance to the samples, the PV parameters of all the samples remain almost unchanged despite the spectral irradiance difference between the sources.

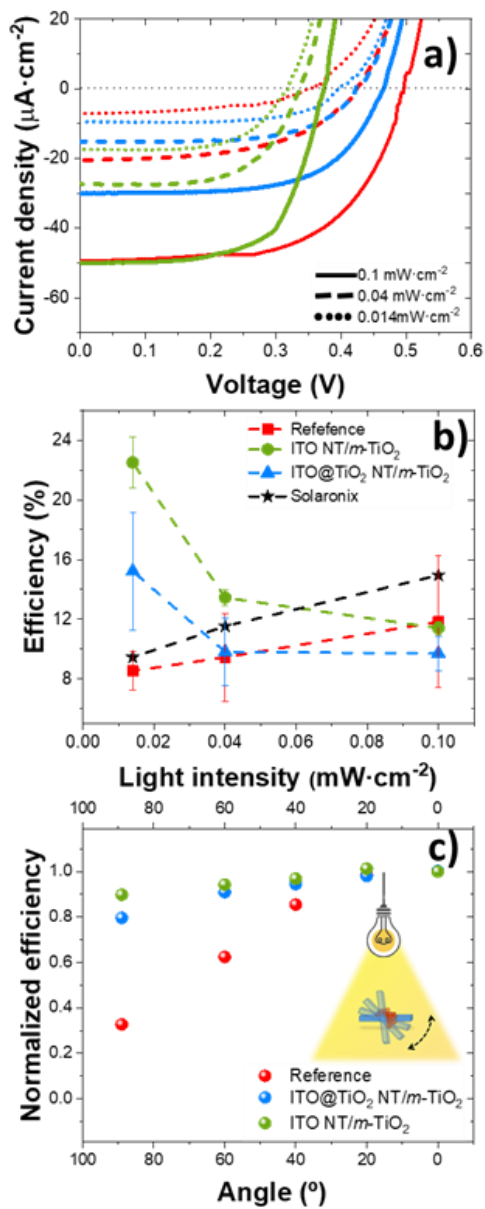

**Figure 5**. Photovoltaic parameters of DSSCs based on ITO NT/*m*-TiO$_2$ and ITO@TiO$_2$ NT/*m*-TiO$_2$, and reference photoanodes under indoor light illumination. a) Current density-voltage curves of the champion cells of DSSCs measured at different light intensities (red, green and blue lines correspond to reference, ITO NT/*m*-TiO$_2$ and ITO@TiO$_2$ NT/*m*-TiO$_2$ respectively). b) Photovoltaic efficiency statistic data obtained from the current density-voltage curves at different light intensities. c) Normalized efficiency obtained from the current density-voltage curves that were reordered at different angles of incidence of light at 0.1 mW·cm$^{-2}$. a), b), c) These data have been acquired under artificial illumination sources

using a cool-white LED (6500 k). The measure has been carried out at reverse scan rate, at 100 mV/s and using a mask of 0.36 cm$^2$

**Table 1**. Photovoltaic parameters statistic of DSSCs based on ITO_NT and ITO@TiO$_2$_NT embedded in a mesoporous scaffold of TiO$_2$, and reference samples obtained from JV curves recorded under indoor light illumination. The photovoltaic parameters have been extracted from current-voltage curves measured at different light intensities in reverse scan (from 1V to -0.2 V) using a mask of 0.36 cm$^2$ and scan rate of 100 mV/s.

|  | Reference | ITONT/ m-TiO$_2$ | ITO@TiO$_2$ NT/m-TiO$_2$ | Reference | ITONT/ m-TiO$_2$ | ITO@TiO$_2$ NT/m-TiO$_2$ | Reference | ITONT/ m-TiO$_2$ | ITO@TiO$_2$ NT/m-TiO$_2$ |
|---|---|---|---|---|---|---|---|---|---|
|  | 0.1 mW·cm$^{-2}$ | | | 0.04 mW·cm$^{-2}$ | | | 0.014 mW·cm$^{-2}$ | | |
| J$_{sc}$ (μA/cm$^2$) | 38.3 ± 15.8 | 47.6 ± 1.4 | 35.8 ± 5.7 | 15.8 ± 6.7 | 27.8 ± 2.3 | 19.6 ± 5.6 | 5.6 ± 2.2 | 17.6 ± 1.5 | 11.8 ± 3.4 |
| V$_{oc}$ (mV) | 515 ± 45 | 364 ± 4 | 411 ± 51 | 460 ± 55 | 337 ± 13 | 358 ± 74 | 392 ± 53 | 286 ± 21 | 318.9 ± 53 |
| FF (%) | 61 ± 3 | 67 ± 1 | 67 ± 4 | 54 ± 1 | 58 ± 5 | 57 ± 3 | 56 ± 6 | 63 ± 5 | 58 ± 11 |
| PCE (%) | 11.9 ± 4.4 | 11.4 ± 0.5 | 9.7 ± 1.1 | 9.5 ± 2.9 | 13.5 ± 0.5 | 9.8 ± 2.3 | 8.5 ± 1.3 | 22.6 ± 1.7 | 15.2 ± 3.9 |

## 2.4 Charge collection and light harvesting mechanisms in multidimensional nanostructured photoelectrodes

As a first step to unravel the main factors involved in the enhanced performance of the nanostructured photoelectrodes, we evaluated the PV performance under a solar simulator at 1-sun of light intensity. Figure S9a shows the best current density-voltage curves (*JV* curves) for each configuration. The average of PV parameters is summarised in Table S1. As can be seen, the highest efficiencies are obtained for the reference samples, closely followed by the ITO nanostructure samples. The decrease in efficiency for the 1D nanostructures is mainly due to the reduced short-circuit photocurrent (J$_{sc}$), although a slight decrease in the open-circuit voltage (V$_{oc}$) is also observed. Focusing on the 1D nanostructure-based DSSCs, the anatase shell in ITO@TiO$_2$ NT/*m*-TiO$_2$ has a positive effect on the V$_{oc}$ and the fill factor (FF) but at the cost of losing the J$_{sc}$. The positive effect of the shell can be explained by a lower rate of electronic recombination and/or conduction band downshift.[29,30] Thus, from highest to lowest efficiency, the devices are ordered as follows: reference > ITO@TiO$_2$ NT/*m*-TiO$_2$ > ITO NT/*m*-TiO$_2$. It should be emphasised that the 1D nanostructures barely influence the dye loading at the photoanode (see Table S1), which is mainly dominated by the thickness of the *m*-TiO$_2$ layer and is slightly higer for the reference device. However, the incident photon to electron conversion efficiency (IPCE) spectra are indeed affected by the nature of the nanostructures, as can be seen in Figure S9b. As expected, the IPCE results agree with the photo-response obtained in the JV curves of Figure S9a.

The electrochemical analysis of the different configurations was carried out using impedance spectroscopy (IS). IS curves were recorded at open-circuit potential over a wide range of light intensities by using a white LED as the light source. This allows the devices to be probed at different positions of the Fermi level in the *m*-TiO$_2$ layer. For DSSCs based on liquid electrolytes, the impedance curve is usually characterized by two signals: the one closest to the origin, that is the high-frequency (HF) signal, is due to charge transfer at the platinum counter-electrode.[45–47] The second signal is the result of the parallel combination of (1) the charge transfer or recombination resistance (R$_{rec}$) and (2) the chemical capacitance of electron accumulation in the semiconductor film (in this case the *m*-TiO$_2$ layer) (C$_\mu$).[48–50] These signals appear as arcs in Nyquist plot (imaginary part vs. real part) (see Figure S10). Both signals are connected by a small 45° transmission line, which is related to the transport resistance (R$_{trans}$) in the metal oxide.[49] Here, the impedance spectra were fitted using the transmission line equivalent circuit developed by Bisquert and coworkers (inset in Figure S10).[51,52] Paying attention to the parameters R$_{rec}$ and C$_\mu$, these evolve with the applied bias according to equations[53,54]

$$R_{rec} \sim exp\left(\frac{-\beta qV}{k_B T}\right) \quad (1)$$

$$C_\mu \sim exp\left(\frac{\alpha qV}{k_B T}\right) \quad (2)$$

where *V* is the applied voltage, $k_B$ is the Boltzmann constant, *q* is the electron unit charge, *T* is the absolute temperature, β is the transfer parameter and α describes the exponential distribution of intra-band states below TiO$_2$ conduction band. [55–57]

Regarding to the voltage dependence of $C_\mu$ (Figure S11), all the structures show the characteristic voltage dependence of DSSCs, where the exponential tendency of $C_\mu$ (Table S2) is consistent with that reported in the literature for mesoporous frameworks ($\alpha$ ~0.15-0.35).[35,43,58] Upon comparing the exponential dependence region of the three devices, a band shift of ~ 50 mV can be observed for the ITO@TiO$_2$ NT/*m*-TiO$_2$ device.[43,59] Therefore, a voltage correction is needed to ensure that the comparison of the recombination resistance of the devices is made with the same electron density in the semiconductor.[52,56,60] This is demonstrated in **Figure 5**a which depicts the $R_{rec}$ vs. corrected voltage data whilst Table S2 reports the $\beta$ obtained for each configuration according to Equation 1. In all cases, the $\beta$ parameter is within the typical values for DSSC ($\beta$ ~0.5-0.8) regardless of the nanostructured photoanode. It is also observed that the recombination resistance follows the trend: reference > ITO@TiO$_2$ NT/*m*-TiO$_2$ > ITO NT/*m*-TiO$_2$, which is in good agreement with the $V_{OC}$ reported in Table S1.

Finally, focusing on the transport resistance in Figure 5b, both nanostructured electrodes show a plateau at low light intensities (i.e. at low values of applied voltage) rather than the classic exponential behaviour of DSSCs (see the reference device).[51] In addition, the DSSC based on ITO NT/*m*-TiO$_2$ electrode has the lowest transport resistance, after the voltage correction. By combining the transport and recombination resistances, the ratio of the diffusion length to the thickness of the active layer can be estimated to provide a qualitative assessment of the charge collection.[52] This ratio can be calculated as follows

$$\frac{L_n}{d} = \sqrt{\frac{R_{rec}}{R_{trans}}} \quad (3)$$

As shown in Figure 5c, this ratio increases sharply for the 1D nanostructured samples at low voltage values. That is, the DSSCs fabricated with the ITO NT/*m*-TiO$_2$ and ITO@TiO$_2$ NT/*m*-TiO$_2$ photoanodes collect the charge more efficiently than the reference sample under low light-intensity conditions.

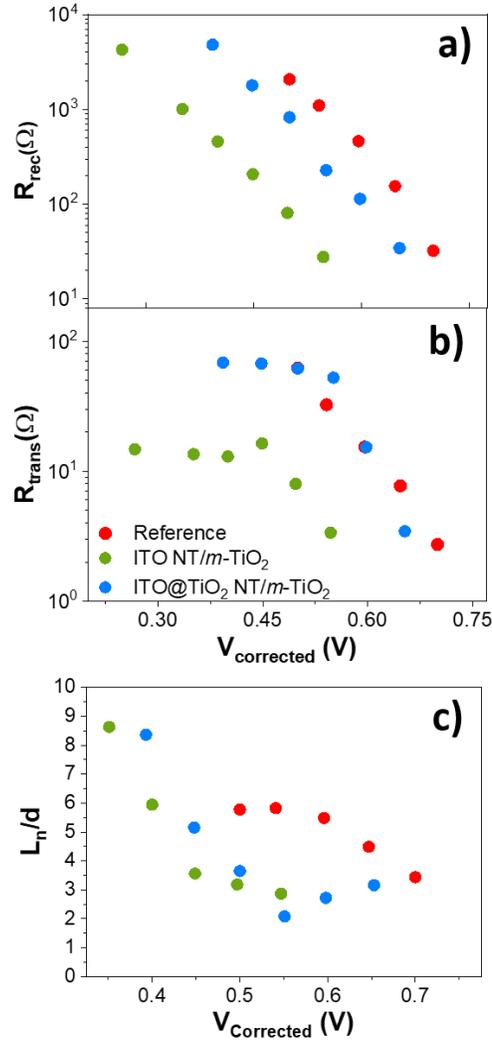

**Figure 4.** Impedance spectroscopy characterization of DSSCs based on ITO NT/*m*-TiO$_2$ and ITO@TiO$_2$ NT/*m*-TiO$_2$, and reference photoanodes. Recombination resistance (a) and transport resistance (b) extracted from Nyquist plots fitting to the equivalent circuit showed in Figure S10. (c) Ratio between small-perturbation diffusion length (Ln) extracted from impedance spectra and the active layer thickness.

Taking into account these results, the enhanced efficiency encountered for the multiscale nanostructured photoanodes at low-light intensity can be explained as follows: Firstly, considering that ITO nanotubes-based devices show higher J$_{sc}$ at low illumination intensity (see Figure S7b and Table 1) compared to the reference. Secondly, a smoother decrease of the photocurrent density with decreasing light intensity is also observed in the case of nanostructured photoanodes (Figure S7b). Thirdly, we also found that the ITO NT/*m*-TiO$_2$ samples maintain a good FF value over the whole illumination range (Figure S7 c). The increase in the J$_{SC}$ for DSSCs based on 1D nanostructures at low- light intensity is explained by, on the one hand, the strong scattering effect observed for the ITO NT and ITO@TiO$_2$ NT shown in Figure 3 and S3, leading to a stronger absorptance in the nanostructured system (Figure S5). This increase in the scattering is a widely reported photon management approach in DSSCs to enhance the photocurrent by increasing the photon path length.[61,62] It should also be noted that the absorption spectrum of the YKP-88 dye is well aligned with the emission of standard indoor incandescent lamps, such as those used in this work (Figure S6). On the other hand, it is also explained by the superior charge collection efficiency of the ITO NT/*m*-TiO$_2$ and ITO@TiO$_2$ NT/*m*-TiO$_2$-based DSSCs (see Figure 5). A proposed schematic representation of the electron transport mechanism and the scattering effect is shown in Scheme 1 and 2c).

## 3. Conclusion

In this work, 1D nanostructured electrodes such as nanotubes of ITO and ITO@TiO$_2$ both embedded in a mesoporous of TiO$_2$ framework have been incorporated into DSSCs as photoanodes. These 1D electrodes are fabricated by a proprietary soft-template method using vacuum and plasma techniques. The 1D nanoelectrodes exhibit pronounced scattering effects, which are visible as a dominant diffuse component of the transmittance and reflectance over a large part of the visible range, in contrast to conventional reference electrodes. Furthermore, this scattering occurs at short wavelengths in the visible range, where most DSSC dyes have their maximum light absorption. ITO NT/*m*-TiO$_2$ and ITO@TiO$_2$ NT/*m*-TiO$_2$ electrodes implemented in DSSCs show the highest efficiency at low-light intensities, such as those recommended for home lighting (0.04-0.014 mW/cm$^2$). The most remarkable PV improvement is found for the ITO NT/*m*-TiO$_2$ samples due to their higher J$_{sc}$ and better FF under these illumination conditions. Specifically, a 166% higher efficiency is achieved compared to conventional photo-anodes. This phenomenon can be explained by the synergy between two effects: (1) a higher fraction of incident light is scattered by 1D nanostructured photoanodes due to their highly diffusive component at short wavelengths in the visible range. Such a feature extends the photon path, which translates to a higher number of excited dye molecules; and (2) they also provide a longer charge diffusion length at low-light intensity, which results in a significantly greater charge collection under these conditions. Thanks to these excellent properties, DSSCs based on 1D nanoelectrodes can retain up to 90% of their initial efficiency under glancing illumination from an indoor light source. In comparison, reference DSSCs lose up to 70% of their efficiency. Therefore, the results obtained in this work suggest that the incorporation of the ITO nanotubes into the photoanodes of DSSCs promotes the regeneration of charges in the electrolyte and in the dye at low light intensities, thus achieving more competitive DSSCs for harvesting indoor illumination.

Author Contributions

The manuscript has been written through the contributions of all authors. L.C.B., R.D., J.R.S.V. and Ana B. did the Conceptualization. The investigation and methodology have been carried out by J.C.S, J.G.R., A.J.R., S.F., Y.K. and L.C.B. All authors have approved the final version of the manuscript.

Acknowledgements

We thank the projects PID2019-109603RA-I00, TED2021-130916B-I00, PID2019-110430GB-C21, and PID2022-143120OB-I00 funded by MCIN/AEI/10.13039/501100011033 and by "ERDF (FEDER) A way of making Europe, Fondos NextgenerationEU and Plan de Recuperación, Transformación y Resiliencia". LCB acknowledges the French embassy in Spain for the grant "123892U" within framework of the proposal scientific call 2022 "Becas doctorales y postdoctorales". Thanks to C.M.D Felipe II, particularly María Inmaculada Carmona Rivera, for providing some tools for preparing counter-electrodes. The project leading to this article has received funding from the EU H2020 program under grant agreement 851929 (ERC Starting Grant 3DScavengers). R.D. acknowledges the European

Research Council (ERC) for funding. This work was funded under the European Union's Horizon 2020 research and innovation programme (grant agreement number 832606; project PISCO).

**Supporting Information**

**Multidimensional nanoarchitectures for improved indoor light harvesting in dye-sensitized solar cells**

*Javier Castillo-Seoane,[a] Lidia Contreras-Bernal,[a,]\* Antonio J. Riquelme,[b,]\* Samuel Fauvel,[b] Yann Kervella,[b] Jorge Gil-Rostra,[a] Gabriel Lozano,[c] Angel Barranco,[a] Renaud Demadrille,[b] Juan R. Sánchez-Valencia[a] and Ana Borrás[a]*

[a.]*Nanotechnology on Surfaces and Plasma Lab., Materials Science Institute of Seville, ICMS, CSIC – US, Avd. Américo Vespucio 49, 41092, Seville (Spain)*

[b.] *Université Grenoble Alpes, CEA, CNRS, INP, IRIG-SyMMES, F-38000 Grenoble, France.*

[c.]*Multifunctional Optical Materials, Materials Science Institute of Seville, ICMS, CSIC – US, Avd. Américo Vespucio 49, 41092, Seville (Spain)*

**Table of contents**

**Experimental section**

*Synthesis of core@shell nanoelectrodes*

ITO nanoelectrodes were fabricated through a multistep procedure involving vacuum and plasma techniques, as previously detailed by our group (See Scheme 2).[27] The process comprised the following sequential steps: (1) Magnetron sputtering to deposit a "seed" layer, utilizing an ITO thin film; (2) Vacuum thermal sublimation to grow metal-free phtalocyanine $H_2Pc$ single crystalline organic nanowires (soft-template) on the ITO thin film; (3) Magnetron sputtering for the conformal deposition of an ITO shell on the organic nanowires; and (4) Air annealing to eliminate the organic core, resulting in ITO-supported nanotubes. Throughout these steps, the fabrication process adhered to mild conditions, maintaining a pressure below $2.0 \times 10^{-2}$ mbar. Substrate temperatures of 350°C were applied for steps (1), (3), and (4), while a temperature of 210°C was employed for step (2). ITO deposition involved radiofrequency magnetron sputtering of a 3" ITO disc target ($SnO_2$ 10 wt% doped $In_2O_3$, Kurt J. Lesker Company), using Ar as a plasma gas, a power of 150 W, and a distance to the samples-holder of 15 cm.

Subsequently, a $TiO_2$ conformal shell was deposited at room temperature using plasma-enhanced chemical vapor deposition (PECVD) with a 2.45 GHz microwave Electron-Cyclotron Resonance (ECR) SLAN-II plasma source. The applied power was 480 W, and O2 served as the plasma gas at a pressure of $1.0 \times 10^{-2}$ mbar. The organometallic precursor for this deposition was titanium (IV) isopropoxide (TTiP, Sigma Aldrich), at 40°C with an $O_2$ bubler as carrier gas. As a result, supported ITO@$TiO_2$ nanotubes.

*Fabrication of dye-sensitized solar cells*

FTO TEC 15 drilled (Solaronix) and ITO (XopGlass, resistance <10 Ω/square; thickness 1.1 mm) have been used as substrates for counter-electrode and working electrode, respectively. Both types of substrates were cleaned with 15 min bath sonication using, successively, Hellmanex© solution, deionized water, acetone and isopropanol. The substrates were then treated in a UV-$O_3$ chamber for 15 min. After that, a platinum solution (Solaronix, Platisol) was spread over the conductive side of the counter-electrodes for subsequent annealing at 400 $^0$C for 5 min. On the other hand, 8 μm thickness of $TiO_2$ colloidal paste (20 nm, GreatCell) was deposited on ITO substrates by screen-printing technique (PET 1500 90/230-40 mesh) for preparing the reference electrodes. The resulting $TiO_2$ layer (referred to as m-$TiO_2$) was sintered at 450 $^0$C for 30 min. In the case of 1D nanostructured working electrodes, the m-TiO$_2$ film was deposited in the same conditions. The active area of electrodes was 0.25 cm$^2$ and thickness was around 8 µm. The mesoporous layer was treated with TiCl$_4$ bath (40 mM aqueous solution) for 20 min at 70ºC. The electrodes were then rinsed with distilled water and absolute ethanol. A mixture of acetonitrile and *tert*-butanol (1:1 % V) containing 0.2 mM of YKP-88 and 2mM of chenodeoxycholic acid was used as dye solution. The working electrodes were immersed into the dye solution overnight in dark and at room temperature under mild agitation. After dye sensitization, the samples were washed with ethanol and dried under air. The liquid-based electrolyte solution was prepared in acetonitrile using iodide/iodine as redox pair: 0.03 M I$_2$ (Sigma-Aldrich), 1M BMII (Solvionic), 0.05 M LiI (Sigma-Aldrich), 0.5 M TBP (Sigma-Aldrich) and 0.1 M GuSCN (Sigma-Aldrich).[36] Later, the working electrode and counter-electrode were sandwiched together using a thin thermoplastic film (Surlyn, 60 µm) at 105 $^0$C under pressure. The assembled device was filled with the electrolyte solution through the hole by vacuum. Finally, the hole was sealed with a piece of thermoplastic film and a coverglass.

*Characterization*

The morphological characterization was carried out by SEM using a Hitachi S4800 microscope at 2kV. The SEM images were treated with ImageJ (free available software) to measure the length and diameter of nanostructures.

TEM images, SAED pattern and EDXS profiles were obtained using a scanning TEM microscope, TALOS F200S from FEI company, working at 200 kV with 0.25 nm resolution. To perform these measurements, the samples were scratched on top of a holey carbon TEM copper grid. For XRD measurements, a Panalytical X'PERT PRO model operating in the θ-2θ configuration with Cu Kα (wavelength of 1.5418 Å) radiation as the excitation source was used.

The UV-vis-NIR spectrophotometer used was an Agilent Technologies Cary 5000 Uv-Vis-NIR equipped with an integrating sphere to acquire absorptance A, diffuse and total transmittance ($T_D$ and $T_T$), diffuse and total reflectance ($R_D$ and $R_T$).

Dye loading has been carried by immersing the photoanodes in a solution of 0.1M aqueous NaOH and THF in a 1:1 %V for 3 hours. After detachment, the dyes are extracted with dichlorotmethane and the analysis is performed by UV-Vis spectroscopy.

The photovoltaic response of DSSCs was studied by recording the current density-voltage curves with a Keithley 2400 at scan rate of 100 mV·s$^{-1}$ in reverse scan and using a mask of 0.36cm$^2$ to delimit the area of the solar cell. For that, different light sources were used: 1) solar simulator (ABET-Sunlite) with AM 1.5G filter at 100 mW·cm$^{-2}$; 2) a cool-white LED of 6500K (lamp 1); 3) warm-white LEDs of 2700K with different luminous fluxes (lamp 2 y lamp 3), and a cool-white fluorescent lamp (lamp 4). The light intensity was calibrated using a Solar Light's Model PMA2100 Dual-Input Data Logging Radiometer with a visible light photometer detector.

For current density-voltage curves at different angles, we used a holder that allows the sample to be tilted with respect to the light source.

Impedance spectroscopy (IS) was carried out at open circuit conditions using a white 6500K LED (lamp 1) as the illumination source. The Fermi level was fixed by the illumination intensity. The IS was measured in the range from $10^6$ to $10^{-1}$ Hz with a perturbation of 20 mV. The frequency response of the DSSCs was analyzed by a response analyzer module (Autolab, PGSTAT302N/FRA2). The IS data was generated by NOVA 2.1.3 software and then analyzed using Z-view equivalent circuit modelling software.

IPCE spectra were measured with a commercial apparatus (Arkeo-Ariadne, Cicci Research s.r.l.) based on a 300 Watt Xenon lamp, from 300 nm to 800 nm with a step of 5 nm.

**Table S1.** Dye loading and photovoltaic parameters statistic of DSSCs based on ITO NT and ITO@TiO2 NT embedded in a mesoporous scaffold of TiO$_2$, and reference samples. The photovoltaic parameters have been extracted from current-voltage curves measured under 1 sun-AM 1.5G illumination in reverse scan (from 1V to -0.2 V) using a mask of 0.36 cm2 and scan rate of 100 mV/s. The highest value obtained is shown in parentheses**.** Photovoltaic parameters statistic of DSSCs based on ITO NT photoanode are also shown.

| Type | $J_{sc}$ (mA/cm²) | $V_{oc}$ (mV) | FF (%) | PCE (%) | Dye loading (mol/cm²) |
|---|---|---|---|---|---|
| ITO NT/*m*-TiO$_2$ | 12.9 (14.1)± 2.2 | 627 (713) ± 34.1 | 48 (52) ± 6.4 | 3.9 (5.3) ±0.9 | 1.1E-07 |
| ITO@TiO$_2$ NT/*m*-TiO$_2$ | 10.8 (11.2) ± 0.5 | 666 (688)± 20.6 | 54 (56) ± 2.5 | 4.1 (4.4) ± 0.3 | 1.1E-07 |
| Reference | 15.4 (15.1) ± 0.8 | 696 (704) ± 5.3 | 52 (57) ± 5 | 5.6 (6) ± 0.4 | 1.4E-07 |

**Table S2.** α and β values extracted from the voltage dependence of the chemical capacitance and recombination resistance (see main text) of DSSCs.

| Type | α | β |
|---|---|---|
| **Reference** | 0.21 | 0.57 |
| **ITO NT/m-TiO$_2$** | 0.26 | 0.46 |
| **ITO@TiO$_2$ NT/m-TiO$_2$** | 0.26 | 0.49 |

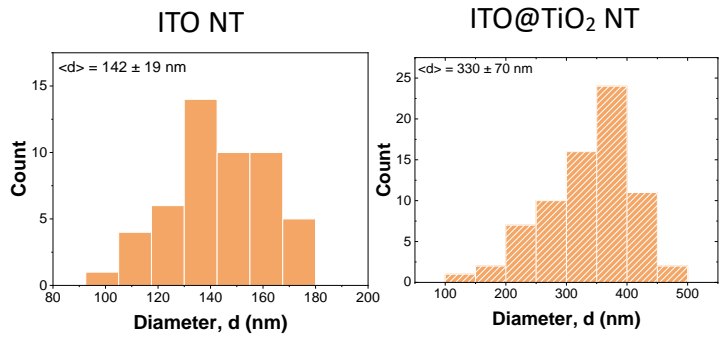

**Figure S1**. Statistical distribution of the widths of ITO and ITO@TiO2 nanotubes obtained from the corresponding SEM images.

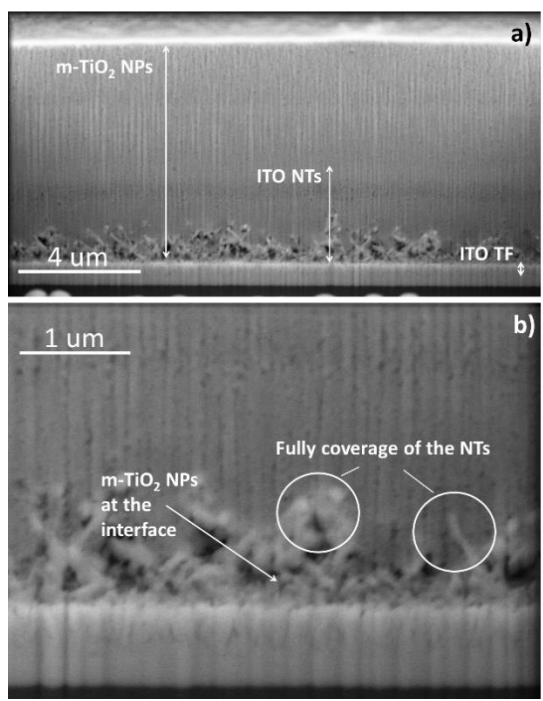

**Figure S2.** FIB Cross section SEM images of ITO NTs + *m*-TiO$_2$ NPs samples showing the homogenous formation of the *m*-TiO$_2$ film with fully coverage of the NTs. Please note that because drifting issues of the microscope scanning acquisition system, the thickness of the ITO thin film interface appears enlarged.

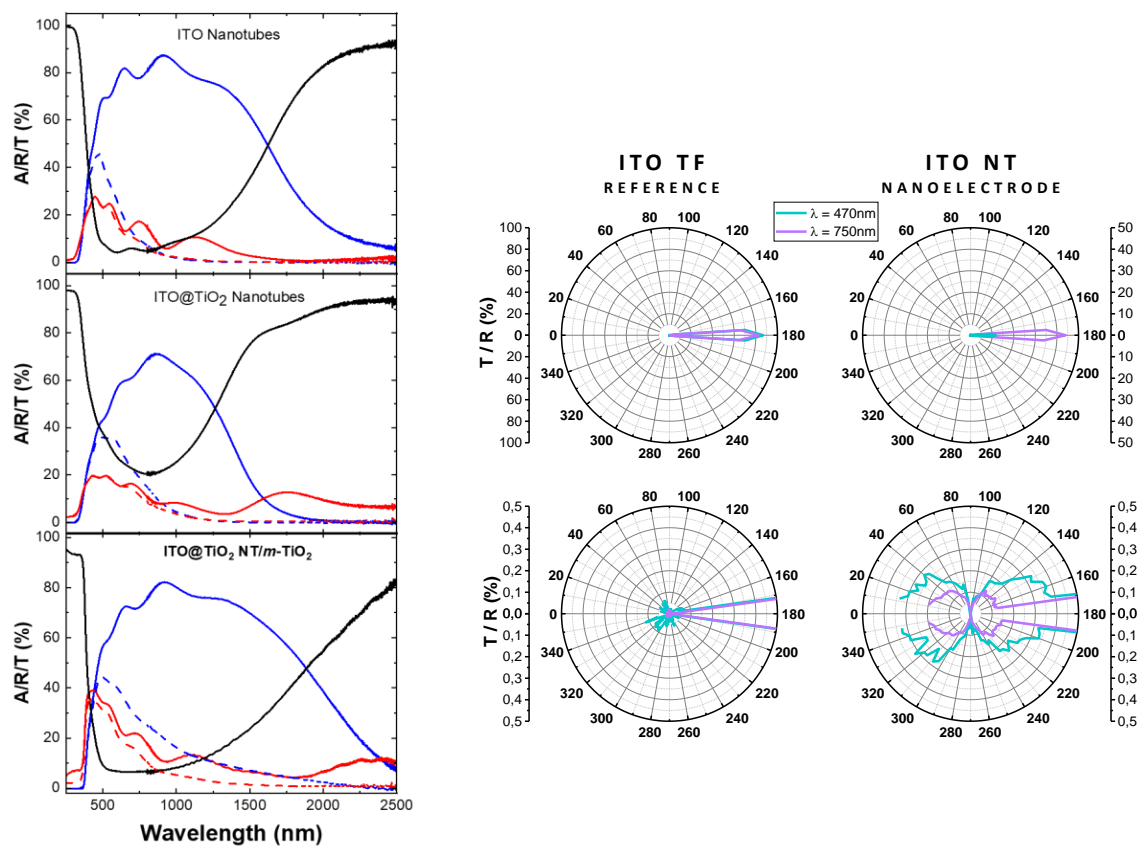

**Figure S3**. (Left) Optical properties of the hetero-nanostructured photoelectrodes. Comparison of transmittance (blue), reflectance (red) and absorptance (black) spectra for diffuse (dash) and total (solid) components for the three nanostructured photoelectrodes comprised of ITO NTs (Top), ITO@TiO$_2$ NTs (Middle) and ITO@TiO$_2$ NTs embedded in *m*-TiO$_2$ film (bottom). (Right) Angular distribution of the light reflected and transmitted of an ITO thin film reference sample and an ITO nanotube sample for λ= 750 nm and 1000 nm.

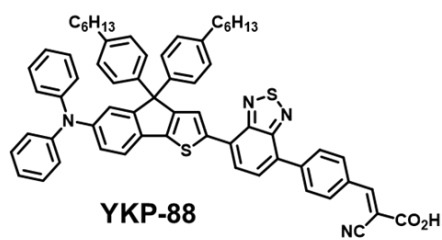

**Figure S4**. chemical structure of the dye.

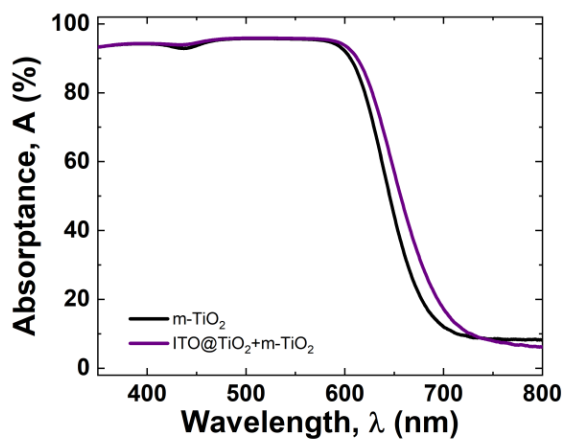

**Figure S5.** The absorptance specta of ITO@TiO2 NT/m-TiO2 electrode and reference sample samples sensitised with the dye YKP-88.

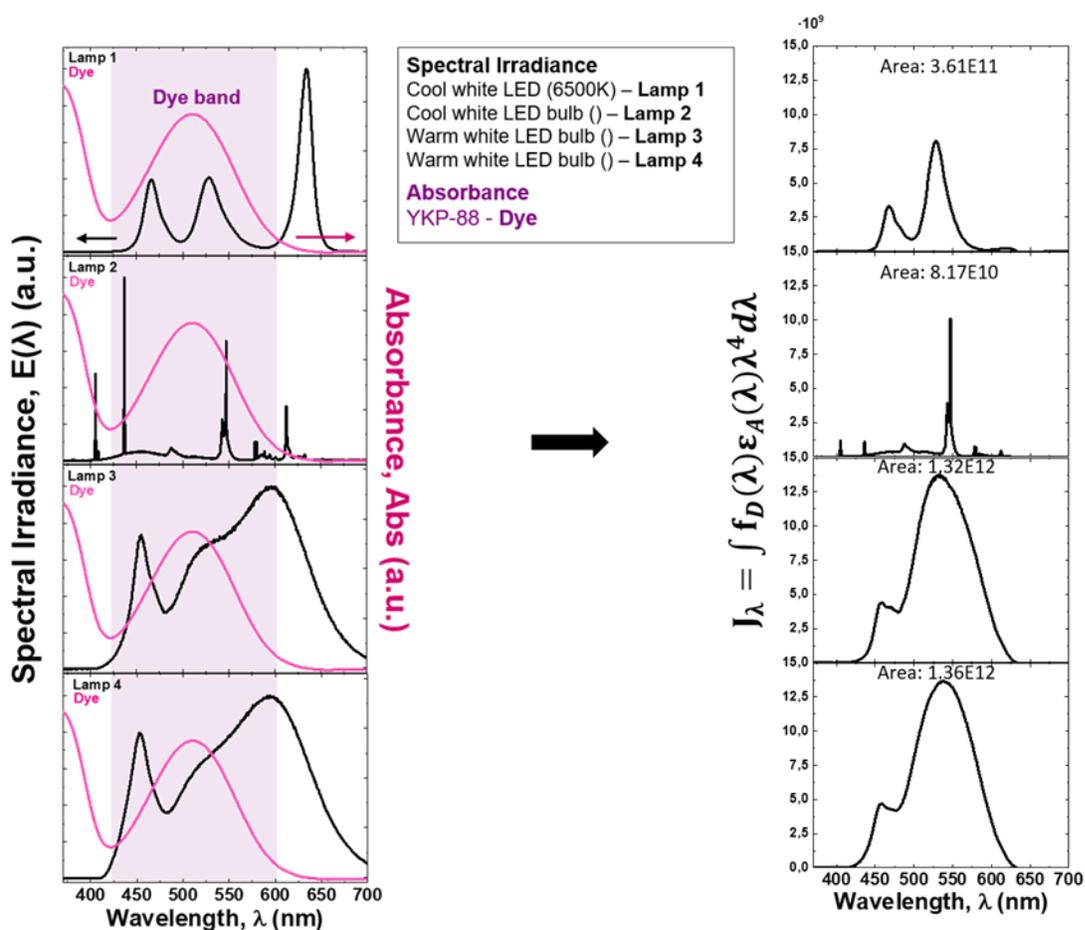

**Figure S6**. (left) Emission spectra from the light sources used for the low intensity analysis and absortion spectra of YKP-88 dye in a solution of 0.1M aqueous NaOH and THF (1:1 % V). (right) Spectral overlap integral.

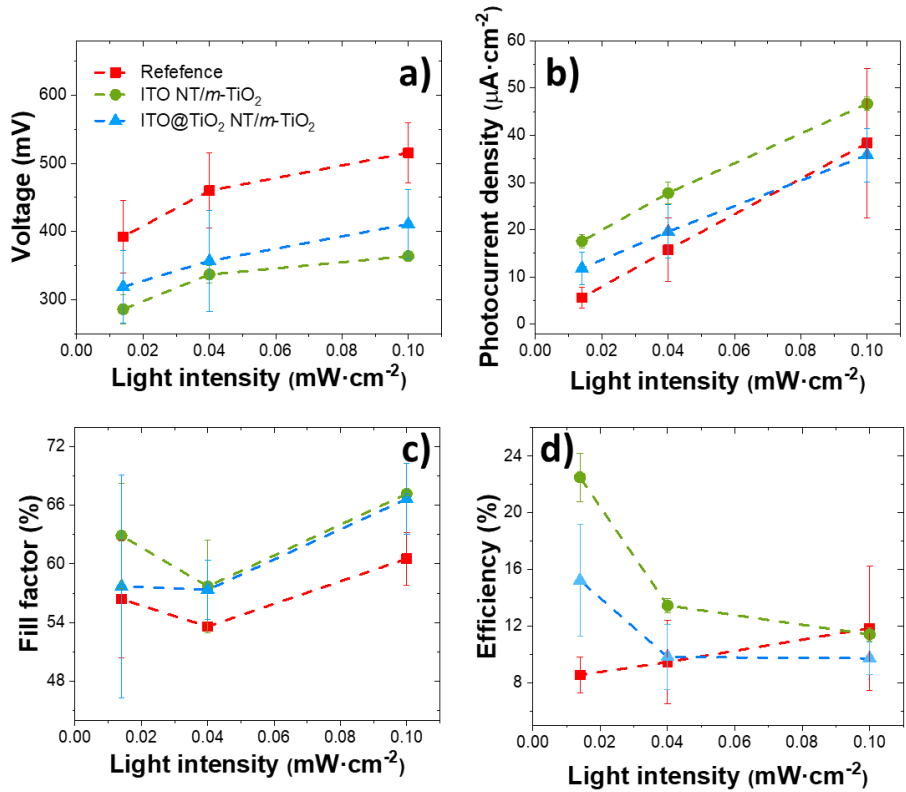

**Figure S7.** Photovoltaic parameters statistics of DSSCs based on ITO NT/*m*-TiO$_2$ and ITO@TiO$_2$ NT/*m*-TiO$_2$, and reference photoanodes under indoor light illumination. These data have been recorded using a white LED (6500 k) at different light intensities. The measure has been carried out at reverse scan, at a rateof 100 mV/s and using a mask of 0.36 cm$^2$.

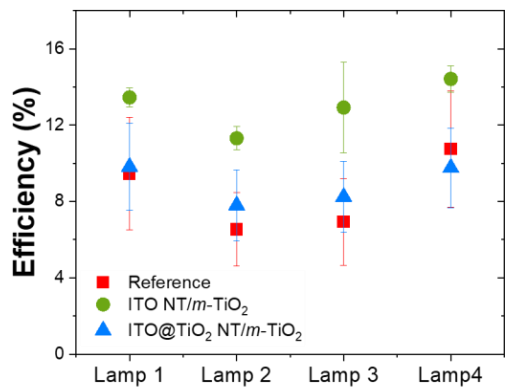

**Figure S8.** Efficiency statistic of DSSCs based on ITO NT/*m*-TiO$_2$ and ITO@TiO$_2$ NT/*m*-TiO$_2$, and reference photoanodes obtained from JV curves recorded under different light bulbs at 0.04 mW/cm$^2$.

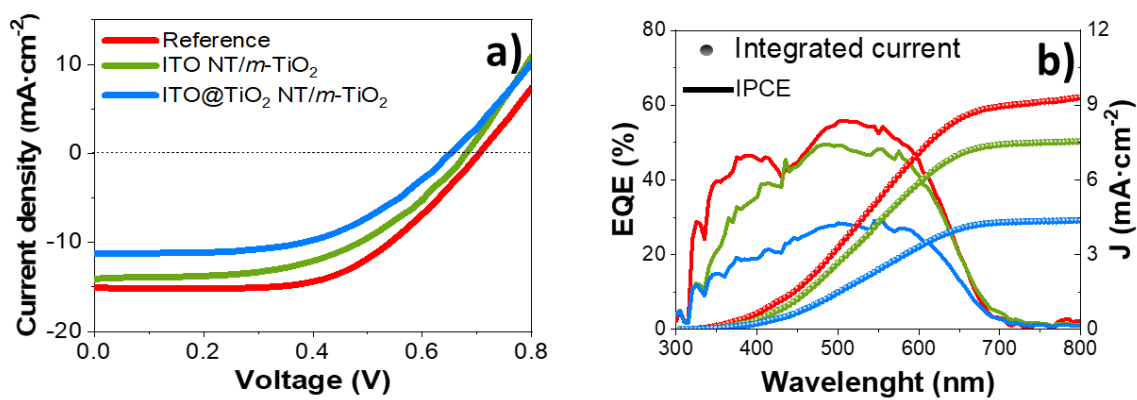

**Figure S9.** (a) Current density-voltage curves of the champion cells. The current density-curves have been measured under 1 sun-AM 1.5G illumination in reverse scan (from 1V to -0.2 V) using a mask of 0.36 cm$^2$ and scan rate of 100 mV/s. (b) The incident photon to electron conversion efficiency (IPCE) spectra of the DSSCs.

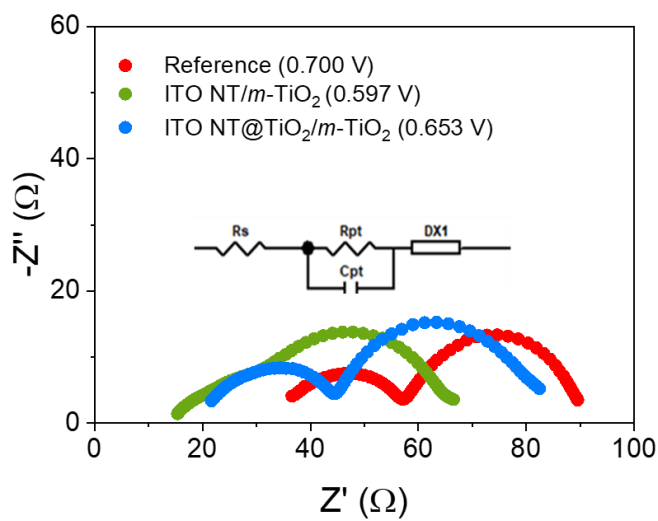

**Figure S10.** Nyquist plot of DSSCs based on ITO NT and ITO@TiO$_2$ NT embedded in a mesoporous scaffold of TiO$_2$, and reference samples. Equivalent circuit used for fitting is included as an inset. The distributed element DX1 corresponds to the distribution line model developed by Bisquert and coworkers (see main text)

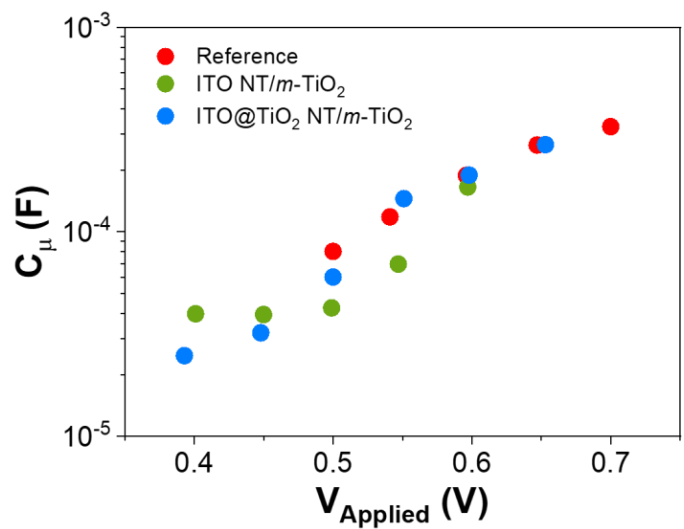

**Figure S11.** Chemical capacitances extracted from fitting the impedance spectra of DSSCs using the equivalent circuit shown in the inset of Figure S4.